\documentstyle[11pt,newpasp,twoside,epsf]{article}

\begin{document}

\title{Theoretical Implications of the 47 Tuc Pulsars}

\author{Frederic A.~Rasio}
\affil{Department of Physics, MIT, Cambridge, MA 02139, USA}

\begin{abstract}
Twenty millisecond radio pulsars have now been observed
in the globular cluster 47 Tuc.
This is by far the largest sample of radio pulsars known in any globular
cluster. 
These recent observations provide a unique opportunity
to re-examine theoretically the formation and evolution of recycled pulsars
in globular clusters. 
\end{abstract}

\section{Introduction}

The observations and properties of the 47~Tuc pulsars are presented by 
Camilo et al.\ (1999) and Freire et al., in these proceedings.
All pulsars are clearly recycled, with pulse periods $P\simeq2-8\,$ms.
At the time of the meeting, accurate timing solutions, including 
positions in the cluster, were known for 14 of the pulsars.
The pulsars can be divided into 3 groups: 7 are single; 8 are in
short-period binaries with orbital periods $P_b<0.5\,$d; 5 are
in wider binaries with $P_b>1\,$d (see \S3).
The measured values of the period derivatives are probaly all
determined predominantly by the pulsar accelerations in the cluster potential
(9 out of 14 pulsars have ${\dot P}<0$).
Under this assumption, Camilo et al.\ (1999) derive a central
density $\rho_c\simeq 4\times10^5\,M_\odot\,{\rm pc}^{-3}$ using
the method developed by Phinney (1993).
This is somewhat larger than previous estimates (Pryor \& Meylan 1993 
give $\rho_c = 1.3\times10^5\,M_\odot\,{\rm pc}^{-3}$).

\section{Radial Distribution}

The radial distribution of the pulsars (Fig.~1) may appear surprising
at first sight. Only 2 pulsars (47 Tuc O and L) are clearly inside 
the cluster core (assumed to have a radius $r_c=12\arcsec$, from
the latest determination by De Marchi et al.\ 1996 based on HST WF/PC
images). One pulsar (47 Tuc F) is near the edge of the core in projection, 
and all others are in the region $r/r_c\simeq1-6$ (the outermost, 47 Tuc C,
has $r/r_c=5.6$). No pulsar is detected in the region $r/r_c\simeq
6-35$, even though this region is well covered by the Parkes beam 
(with a half-power diameter of about $14\arcmin$ at $20\,$cm).

The roughly flat histogram in $r$ (left panel of Fig.~1) suggests
a deprojected 3D number density of pulsars $n(r)\propto r^{-2}$
out to $r\simeq 6\,r_c$, while the absence of pulsars outside that
region indicates a much steeper density profile for $r>6\,r_c$.
This is actually consistent with a thermally relaxed radial distribution in the 
cluster, not too different from that of other stellar components
such as red giants (right panel of Fig.~1). This is to be expected
since the central relaxation time in 47~Tuc, $t_{rc}\sim10^8\,$yr, is much
shorter than typical characteristic ages of millisecond pulsars
($t_c\ga10^9\,$yr). A very similar type of radial distribution has
been observed for the 
7 single millisecond pulsars near the center of M15 (see Phinney
1993 for a detailed theoretical analysis).
The small eccentricities of
the binaries also indicate that the observed pulsars have not been
perturbed by strong interactions with other cluster stars or binaries,
consistent with their present positions in the cluster
(see Rasio \& Heggie 1995). The concentration of wider binaries
(47~Tuc E, H, and ~Q) well outside the core, at $r\ga3\,r_c$, is
also consistent with the smaller collision time in this region
(the collision time for these binaries would have been $\sim10^9\,$yr
inside the core, but it increases to $\sim10^{10}\,$yr at $r=3\,r_c$).
However, for the somewhat large eccentricity of 47~Tuc H, $e=0.07$,
to be explained by perturbations from other cluster stars, the
results of Rasio \& Heggie (1995) show that this
pulsar must have spent a significant fraction of its life inside
the cluster core, in spite of its present location at $r=4\,r_c$.

\begin{figure}
\plottwo{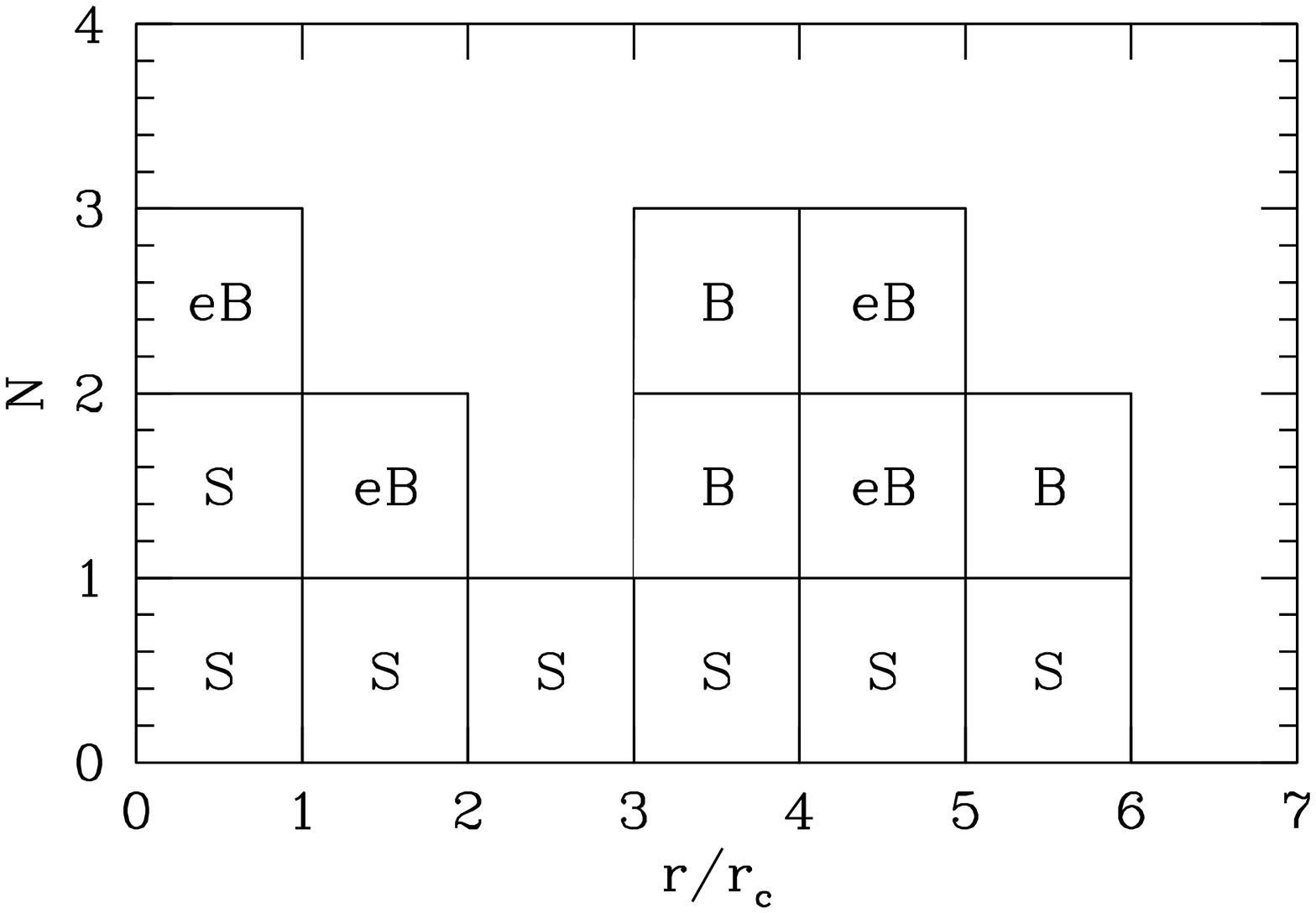}{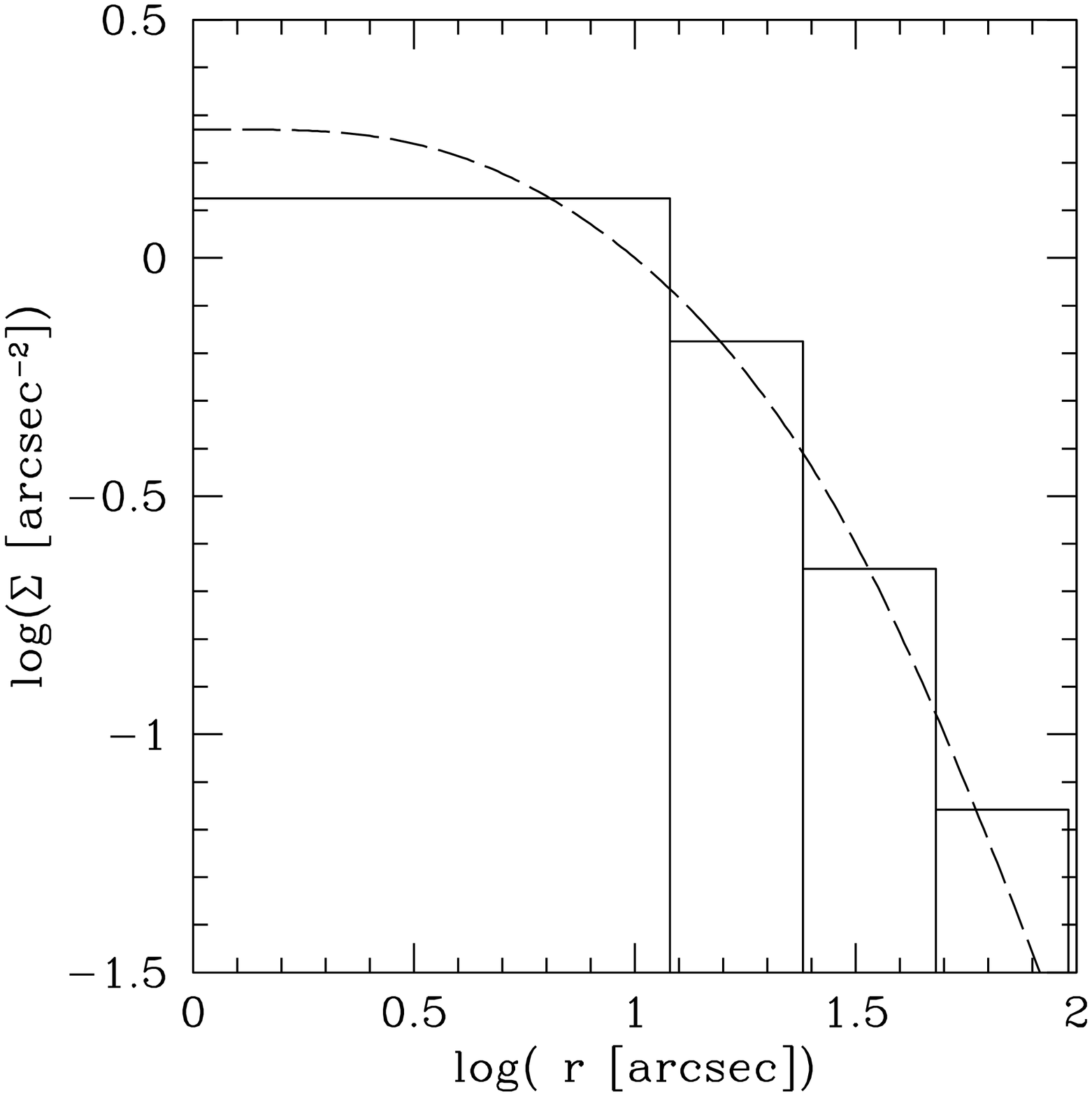}
\caption{Projected radial distribution of the 47 Tuc pulsars. On the left is 
the histogram of
the number of pulsars in each radial bin. The radius (distance to the
cluster center) is given in units of the core radius $r_c=12\arcsec$ (from
the HST WF/PC images of De Marchi et al.\ 1996). The labels
have the following meaning: S denotes a single pulsar, eB an ``eclipsing binary''
($P_b<1\,$d), and B a ``normal binary'' ($P_b>1\,$d). On the right, the surface 
density of pulsars (histogram) 
is compared to that of giants (dashed line; from Fig.~2 of De Marchi et al.\ 1996).
Pulsars were grouped into one core bin ($r/r_c=0-1$) and 3 logarithmically spaced
outer bins ($r/r_c=1-2$, $2-4$, and $4-8$). The surface density of pulsars was
rescaled to match that of giants at $\log(r[\arcsec])=1.55$ (center of the third bin).}
\end{figure}

\begin{figure}
\plotone{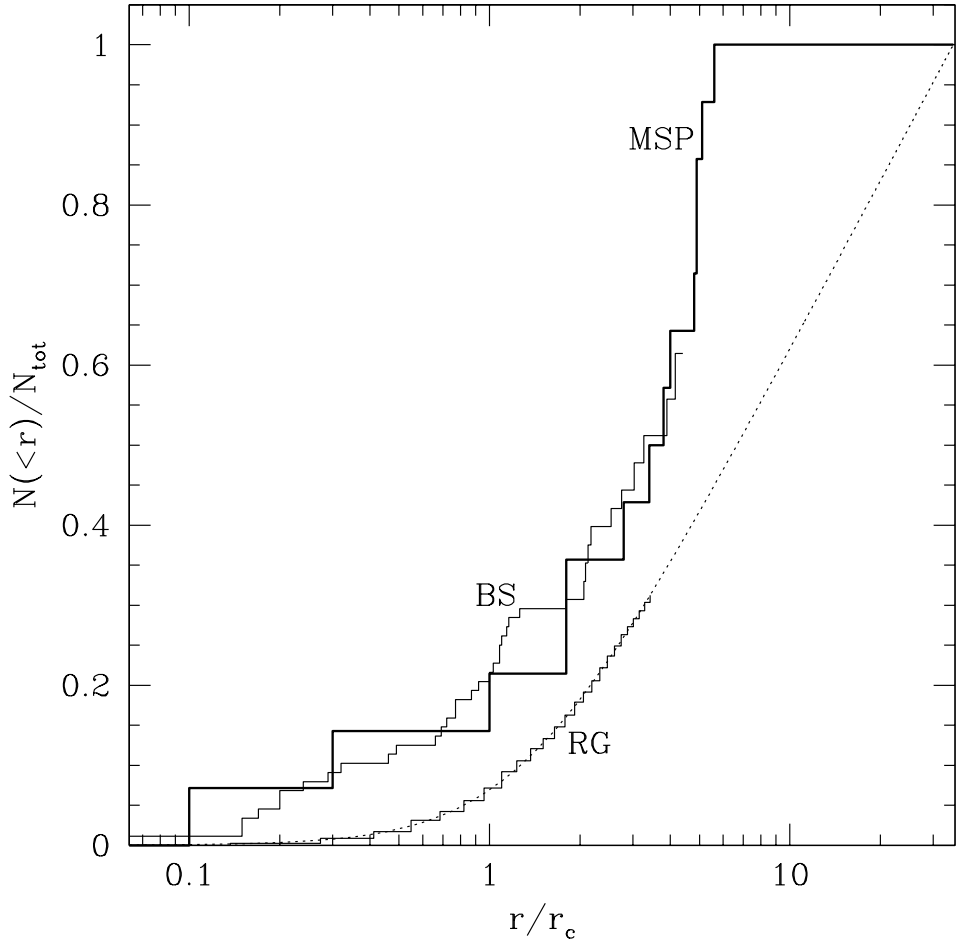}
\caption{Cumulative radial distribution of the 47 Tuc pulsars (MSP, thick
histogram), compared to that of
red giants (RG) and blue stragglers (BS). The RG distribution was calculated by integrating
the surface density profile given in Table~1 of De Marchi et al.\ (1996). The dotted
line is the best fit to a modified Hubble law (which is analytic and closely
approximates the inner regions of high-concentration King models; see, e.g.,
Binney \& Tremaine 1987), extended to $r=35\,r_c$ (near the edge the Parkes
beam at $20\,$cm). Note that fitting the cumulative distribution leads to a
tighter constraint on $r_c$ than fitting the surface density (as done by De Marchi
et al.). The best-fit model shown here gives a core radius about 30\% larger
than that quoted by De Marchi et al.\ (although we still give $r$ in units
of $r_c=12\arcsec$, as in Fig.~1). The BS distribution
was obtained by combining the data sets of Paresce et al.\ (1991, HST FOC) and
Guhathakurta et al.\ (1992, HST PC). Star HST-12 of Paresce et al.\ was adopted as
the cluster center. The surface density at $r\ga2\,r_c$ was corrected for incomplete
radial annuli in the HST PC image (as in De Marchi et al.\ 1996). 
The BS distribution was normalized to match that of the millisecond
pulsars at $r\simeq4.2\,r_c$ (extent of the HST PC image).}
\end{figure}

In a thermally relaxed stellar system, equipartition of
energy leads to a more centrally concentrated spatial distribution for
more massive stars. Since
the main-sequence turnoff mass in the cluster ($m_{to}\simeq 0.85\,M_\odot$
for 47~Tuc) is  a factor
of $\sim1.5-2$ lower than the masses of neutron star systems (including the
companion masses for binaries), we would expect the 
radial distribution of pulsars to be more centrally concentrated than
that of giants. This can be seen most convincingly by comparing integrated
surface density profiles (Fig.~2). HST observations of the central region 
of 47 Tuc have revealed a large number of centrally concentrated,
bright blue stragglers (Paresce et al.\ 1991;
Guhathakurta et al.\ 1992), which are thought
to be formed through collisions and mergers of main-sequence stars in the
dense cluster core (Lombardi et al.\ 1996; Sills et al.\ 1999).
These blue stragglers are main-sequence stars well above the turnoff, 
with theoretically derived masses in the range $m_{bs}\simeq 1.4-1.7\,M_\odot$ 
(one has a directly measured mass of $1.7\pm0.4\,M_\odot$; see Shara et al.\ 1997), 
i.e., very similar to the masses of neutron star systems. Indeed, we see in
Fig.~2 that the radial distributions of blue stragglers and millisecond
pulsars in 47~Tuc appear strikingly similar.

\section{Formation of Recycled Pulsars in Short-Period Binaries}

The binary properties of the 47~Tuc pulsars are quite surprising. While 
7 pulsars are single, the majority are in short-period binaries. Most of
the binaries (8 out of 13) have properties similar to those of the rare ``eclipsing
binary pulsars'' (5 do indeed show evidence for radio eclipses) seen in the 
Galactic disk population (see Nice, these proceedings, for a review).
These systems have extremely short orbital periods, $P_b\sim1-10\,$hr, 
circular orbits, and
very low-mass companions, with $m_2\sin i\sim 0.01-0.1\,M_\odot$.
The remaining 5 binaries have properties more similar to those of the bulk
disk population, with nearly-circular orbits, periods $P_b\sim1-3\,$d (near
the short-period end of the distribution for binary millisecond pulsars
in the disk) and companions of mass $m_2\sin i\simeq 0.2\,M_\odot$
(presumed to be white dwarfs, hereafter WD).
The large inferred total population of recycled
pulsars in 47 Tuc ($\sim10^3$, see Camilo et al.\ 1999), 
as well as the high central density of the cluster, suggest that
dynamical interactions must play a dominant role in the formation of
these systems. However, the two dynamical formation scenarios traditionally
invoked for the production of recycled pulsars in globular clusters
run into many difficulties.

Scenarios based on {\em tidal capture\/}
of low-mass main-sequence stars (MS) by
neutron stars (NS), followed by accretion 
and recycling of the NS
during a stable mass-transfer phase, fell out of favor many years ago. 
Serious problems were pointed out about the tidal capture process 
itself (which, because of strong nonlinearities in the regime relevant to
globular clusters, is far more likely to result in a merger than in the formation 
of a detached binary; see, e.g., Kumar \& Goodman 1996; McMillan et al.\ 1990;
Rasio \& Shapiro 1991; Ray et al.\ 1987). 
Moreover, the basic predictions of tidal capture
scenarios are at odds with many observations of binaries in
clusters (Bailyn 1995; Johnston et al.\ 1992; Shara et al.\ 1996).
It is likely that 
``tidal-capture binaries'' are either never formed, or contribute
negligibly to the production of recycled pulsars.
Verbunt (1987) proposed that collisions between NS and red giants
might produce directly NS--WD binaries with ultra-short 
periods, but
detailed hydrodynamic simulations later showed that this does not
occur (Rasio \& Shapiro 1991).

The viability of tidal capture and two-body collision
scenarios has become less relevant with
the realization over the last 10 years that globular clusters contain
dynamically significant populations of {\em primordial binaries\/} (Hut et al.\ 1992).
Neutron stars can then acquire MS companions through {\em exchange interactions\/}
with these primordial binaries. Because of its large cross section, this 
process dominates over any kind of two-body interaction even for low
primordial binary fractions (Heggie et al.\ 1996; Leonard 1989; 
Sigurdsson \& Phinney 1995). 
In contrast to tidal capture, exchange reactions with
hard primordial binaries (with semimajor axes $a\sim0.1-1\,$AU) 
can form naturally the wide
binary millisecond pulsars seen in some low-density globular clusters
(such as PSR B1310$+$18, with $P_b=256\,$d, in M53, which has the lowest
central density, $\rho_c\sim10^3\,M_\odot\,{\rm pc}^{-3}$, of any globular
cluster with observed radio pulsars; see, e.g., Phinney 1996). 
When the newly acquired MS companion, of mass 
$\la1\,M_\odot$, evolves up the giant branch, the orbit circularizes
and a period of {\em stable\/} mass transfer
begins, during which the NS is recycled
(see, e.g., Rappaport et al.\ 1995). The resulting 
NS--WD binaries have orbital periods in the range $P_b\sim1-10^3\,$d.
However, this scenario
cannot explain the formation of recycled pulsars in binaries with
periods shorter than $\sim 1\,$d. To obtain such short periods,
the initial primordial binary must be extremely hard, with $a \la 0.01\,$AU,
but the recoil velocity of the system following the exchange interaction
would then almost certainly exceed the escape speed from the shallow
cluster potential ($v_e\simeq 60\,{\rm km}\,{\rm s}^{-1}$ for 47~Tuc).

One can get around this problem by considering more carefully
the stability of mass transfer in NS--MS
binaries formed through exchange interactions.
While all MS stars in the cluster {\em today\/} have masses $\la1\,M_\odot$,
the rate of exchange interactions may very well have peaked at a time
when significantly more massive MS stars were still present.
Indeed, the NS and the most massive primordial binaries will
undergo mass segregation and concentrate in the cluster core on a time scale
comparable to the initial half-mass relaxation time $t_{rh}$. For a dense
cluster like 47 Tuc, we expect $t_{rh}\simeq 10^9\,$yr (slightly lower 
than the present value),
which is comparable to the MS lifetime of a $\simeq 2-3\,M_\odot$ star.
If the majority of NS acquired these more massive companions, a
drastically different evolution follows. Indeed, in this case, 
when the MS star evolves and fills its Roche lobe, the
mass transfer is dynamically unstable and leads to a common-envelope (CE) phase.
The emerging binary will have a low-mass WD in
a short-period, circular orbit around the NS.
This simple idea is at the basis of the dynamical scenario developed
recently by Rasio, Pfahl, \& Rappaport (1999). They find that a
significant fraction of the NS--WD binaries emerging
from the CE phase undergo further evolution
during a period of stable mass transfer driven by gravitational radiation 
and tidal heating of the companion. The properties of the final binaries
match very well the observed properties of very short-period (eclipsing) 
binaries in 47~Tuc. 
A similar scenario, but starting from tidal capture binaries and applied to 
X-ray sources in globular clusters, was discussed by Bailyn \& Grindlay (1987).
The possibility of forming 
intermediate-mass binaries through exchange interactions was mentioned
by Davies \& Hansen (1998), who pointed out that NS retention 
in globular clusters may
also require that the NS be born in massive binaries. Among
eclipsing pulsars in the disk, at least one system (PSR J2050$-$0827) 
is likely to have had an intermediate-mass binary progenitor, 
given its very low transverse velocity (Stappers et al.\ 1998).

\acknowledgments
This work was supported by NSF Grant AST-9618116 and NASA ATP 
Grant NAG5-8460, and by a Sloan Research Fellowship.

\end{document}